\documentclass[secnumarabic,12pt,tightenlines,eqsecnum,floats,aps,amsmath,amssymb,nofootinbib,superscriptaddress,showpacs,prd]{revtex4-1}

\usepackage{amsmath,amsthm,amssymb,amsfonts}
\usepackage{graphicx}
\usepackage{supertabular}
\usepackage{mathcomp}
\usepackage{float}
\usepackage{units}
 \usepackage{hyperref}
\usepackage{bm}

\numberwithin{equation}{section}

\begin{document}

\title{Band structure in the polymer quantization of the harmonic oscillator}

\author{J. Fernando \surname{Barbero G.}}
\email[]{fbarbero@iem.cfmac.csic.es}
\affiliation{Instituto de
Estructura de la Materia, CSIC, Serrano 123, 28006 Madrid, Spain}
\affiliation{Grupo de Teor\'{\i}as de Campos y F\'{\i}sica Estad\'{\i}stica, Instituto Universitario Gregorio Mill\'an
Barbany, Universidad Carlos III de Madrid, Unidad Asociada al IEM-CSIC.}

\author{Jorge \surname{Prieto}}
\email[]{jorgeprietoarranz@gmail.com}
\affiliation{Instituto Gregorio
Mill\'an, Grupo de Modelizaci\'on y Simulaci\'on Num\'erica,
Universidad Carlos III de Madrid, Avda. de la Universidad 30, 28911
Legan\'es, Spain}

\author{Eduardo J. \surname{S. Villase\~nor}}
\email[]{ejsanche@math.uc3m.es}
\affiliation{Instituto Gregorio
Mill\'an, Grupo de Modelizaci\'on y Simulaci\'on Num\'erica,
Universidad Carlos III de Madrid, Avda. de la Universidad 30, 28911
Legan\'es, Spain}
\affiliation{Grupo de Teor\'{\i}as de Campos y F\'{\i}sica Estad\'{\i}stica, Instituto Universitario Gregorio Mill\'an Barbany, Universidad Carlos III de Madrid, Unidad Asociada al IEM-CSIC.}

\begin{abstract}
We discuss the detailed structure of the spectrum of the Hamiltonian for the polymerized harmonic oscillator and compare it with the spectrum in the standard quantization. As we will see the non-separability of the Hilbert space implies that the \textit{point} spectrum consists of bands similar to the ones appearing in the treatment of periodic potentials. This feature of the spectrum of the polymeric harmonic oscillator may be relevant for the discussion of the polymer quantization of the scalar field and may have interesting consequences for the statistical mechanics of these models.
\end{abstract}

\maketitle

\section{Introduction}{\label{intro}}

The polymer quantization of the harmonic oscillator has been considered as a toy model to gain a deeper understanding about some features of loop quantum gravity (LQG) \cite{AFW,CVZ,CVZ1}. It has also been used to discuss several relevant issues in loop quantum cosmology (LQC) \cite{AshtekarSingh,Bojowald} and may be useful to study field theories such as the scalar field \cite{L1,L2,L3,DDL,Husain1}. One of the consequences of the use of the more exotic, \textit{non-separable}, Hilbert spaces characteristic of this approach is that the uniqueness theorem of Stone-von Neumann --in the case of quantum mechanical systems with a finite number of degrees of freedom-- can be sidestepped and it is then possible to work with non-standard representations of the Weyl algebra (see \cite{Halvorson} for the use of the polymer approach to discuss conceptual issues in quantum mechanics).

The available literature on this subject discusses some features of the spectrum of one of the infinitely many possible generalizations of the standard harmonic oscillator Hamiltonian that can be written in the new setting. The particular choice made in the literature gives a well know differential equation --the Mathieu equation-- for the eigenvalue problem and has led to the conclusion that the spectrum of the polymerized Hamiltonian is very similar to the standard one in quantum mechanics in the limit where the characteristic length parameter $q_0\ell$, that controls the quantum polymer harmonic oscillator Hamiltonian, becomes very small \cite{CVZ,CVZ1,CV}. (Here, $\ell$ has dimension of length and $q_0$ is dimensionless.) There is, however, a disturbing issue that should be addressed. If the spectrum is a pure point one consisting of a countable set of non-degenerate eigenvalues (close to those of the standard quantized harmonic oscillator) then something must be missing because the countable set of eigenvectors cannot provide an orthonormal basis for the non-separable Hilbert space of the polymerized harmonic oscillator (any such orthonormal basis should be uncountable). As we show in the paper the actual spectrum of the Hamiltonian operator for the polymerized harmonic oscillator consists of an uncountable number of eigenvalues grouped in \textit{bands}, very much like those appearing in the study of periodic potentials in standard quantum mechanics. The main difference between the spectrum of the polymerized harmonic oscillator and that of a standard periodic potential is that, in the former case, the points in the spectrum are actual eigenvalues (i.e. they belong to the point spectrum) associated with \textit{normalizable} states whereas for the standard quantization of periodic potentials the bands correspond to the \textit{continuous} spectrum. As a consequence of this, although some parts of the spectrum of the polymerized harmonic oscillator (in particular the lowest eigenvalues), resemble the lowest lying part of the standard harmonic oscillator spectrum --when the characteristic length $q_0\ell$ tends to zero-- some crucial differences arise. These are associated, in particular, with the fact that in the ``limit'' $q_0\rightarrow 0$  the eigenvalues become infinitely degenerate. This has some far reaching consequences, in particular, for the statistical mechanics of these models because the standard definition for both the statistical entropy in the microcanonical ensemble and the partition function in the canonical one fail to be well defined. It may also be relevant for the polymer quantization of the scalar fields --that relies heavily upon the quantization of the harmonic oscillator.

The paper is organized as follows. After this introduction we briefly review in section \ref{Hilbertspaces} the main features of the Hilbert space of almost periodic functions on the real line  $AP(\mathbb{R},\mathbb{C})$ or, equivalently, $L^2(\bm{b}\mathbb{R},\nu_{\bm{b}\mathbb{R}})$. We will emphasize, in particular, the fact that it is isomorphic to the Hilbert space $\ell^2(\mathbb{R})$ (the isomorphism realized in concrete terms by the Fourier-Bohr transform). We will quickly discuss in section \ref{PolyWeyl} the position and momentum representations of the Weyl algebra in this framework. We will then present different ways of writing suitable Hamiltonians that mimic the standard Hamiltonian for the ordinary harmonic oscillator in the ``limit'' in which certain parameter --that has to be introduced to approximate one of the quadratic pieces of the ordinary Hamiltonian-- goes to zero. In section \ref{spectra} we study in detail the spectrum of the resulting Hamiltonians in the non-separable space of polymerized quantum mechanics. By relying on mathematical theorems developed since the sixties we will show that the spectra of these Hamiltonians can be obtained by considering the auxiliary problem of obtaining the spectrum of a Hamiltonian with a periodic potential in the separable Hilbert space $L^2(\mathbb{R},\mu_{\mathbb{R}})$, where $\mu_{\mathbb{R}}$ is the Lebesgue measure. This is straightforward because the Bloch theorem (or the Floquet theory for differential equations with periodic coefficients) provides a complete solution. The main features of the band spectrum for periodic potentials have been extensively studied \cite{ReedSimon} and many rigorous, and even completely solvable, examples have been discussed in the literature. We briefly discuss some concrete examples of Hamiltonians for the polymerized harmonic oscillator in section \ref{examples}. In addition to the popular Hamiltonian that leads to the Mathieu equation for the computation of the energy eigenvalues, we will also comment on another natural choice for the potential: a periodic extension of a purely quadratic potential defined on a symmetric interval about the origin with a periodicity controlled by the characteristic length parameter $q_0\ell$. We will also comment briefly on the Lam\'e potentials in order to argue that the number of bands in the spectrum can actually be finite. The main reason to do this is to emphasize the enormous ambiguity present in the problem. We show in section \ref{statistical} how the structure of the spectrum makes it difficult to define the partition function for these models and study their statistical mechanics \cite{Morales1}, at least if one relies on standard energy-based ensembles and does not introduce extra conditions to eliminate unwanted parts of the spectrum. We end the paper in section \ref{conclusions} with our conclusions and comments regarding, in particular, the relevance of the banded structure of the spectrum for the physical applications of these models.

\section{Polymer Hilbert spaces}{\label{Hilbertspaces}}

The Hilbert spaces $L^2(\bm{b}\mathbb{R},\nu_{\bm{b}\mathbb{R}})$ and $AP(\mathbb{R},\mathbb{C})$, that will be use in the paper,  are unitarily isomorphic to the non-separable complex Hilbert space
$$
\ell^2(\mathbb{R}):=\{\Psi:\mathbb{R}\rightarrow \mathbb{C}\,:\, \sum_{x\in \mathbb{R}} |\Psi(x)|^2<\infty\}
$$
with scalar product
$$
\langle\Psi_1,\Psi_2\rangle_{\ell^2(\mathbb{R})}=\sum_{x\in \mathbb{R}}\overline{\Psi_1(x)}\Psi_2(x)\,.
$$
Notice that, as the sums extend over the whole real line, the set $\{x\in \mathbb{R}:\Psi(x)\neq 0\}$ must be a finite or countable subset of $\mathbb{R}$ for each $\Psi\in \ell^2(\mathbb{R})$.

For every $x_0\in \mathbb{R}$, let us denote by $\delta_{x_0}\in \ell^2(\mathbb{R})$ the characteristic function of the set $\{x_0\}\subset \mathbb{R}$, i.e.
$$
\delta_{x_0}(x):=\left\{\begin{array}{lcl} 1 &\mathrm{ if }& x=x_0\\ 0 &\mathrm{ if }& x\neq x_0\end{array}\right.
$$
The family $\{\delta_x:x\in \mathbb{R}\}$ constitutes an ortonormal basis of $\ell^2(\mathbb{R})$ and, hence, any $\Psi\in \ell^2(\mathbb{R})$ can be written in the form
$$
\Psi=\sum_{x\in \mathbb{R}} \langle \delta_x , \Psi\rangle_{\ell^2(\mathbb{R})}\,\delta_x =\sum_{x\in \mathbb{R}} \Psi(x)\,\delta_x\,.
$$

Let us denote by  $\nu_{\bm{b}\mathbb{R}}$ the probabilistic Haar measure on $\bm{b}\mathbb{R}$ (the Bohr compactification of the real line) and consider the Hilbert space $L^2(\bm{b}\mathbb{R},\nu_{\bm{b}\mathbb{R}})$ with scalar product
$$
\langle\psi_1,\psi_2\rangle_{\bm{b}\mathbb{R}}=\int_{\bm{b}\mathbb{R}}\overline{\psi_1}\psi_2 \nu_{\bm{b}\mathbb{R}}\,.
$$
The space  $L^2(\bm{b}\mathbb{R},\nu_{\bm{b}\mathbb{R}})$ is isomorphic to the Hilbert space $AP(\mathbb{R},\mathbb{C})$ of the (complex) Besicovitch almost periodic functions on $\mathbb{R}$, with  scalar product given by
$$
\langle\psi_1,\psi_2\rangle_{AP}=\lim_{T\rightarrow \infty}\frac{1}{2T}\int_{-T}^T\overline{\psi_1(y)}\psi_2(y)\,\mathrm{d}y\,.
$$
The Fourier-Bohr transform
$$\mathfrak{B}:L^2(\bm{b}\mathbb{R},\nu_{\bm{b}\mathbb{R}})\rightarrow \ell^2(\mathbb{R})$$
is an unitary isomorphism between $L^2(\bm{b}\mathbb{R},\nu_{\bm{b}\mathbb{R}})$ and $\ell^2(\mathbb{R})$. Explicitly, for every $\psi\in L^2(\bm{b}\mathbb{R},\nu_{\bm{b}\mathbb{R}})$  we have
$$
\mathfrak{B}(\psi)(x):=\langle \chi_x,\psi \rangle_{\bm{b}\mathbb{R}}=\int_{\bm{b}\mathbb{R}}\chi_{-x} \psi \nu_{\bm{b}\mathbb{R}}
$$
where, for every $x\in \mathbb{R}$,  $\chi_x$ denotes the continuous character of $\bm{b}\mathbb{R}$ such that $\chi_{x}(\imath(y))=e^{i xy}=:\phi_x(y)$, for all $y\in \mathbb{R}$ and $\imath:\mathbb{R}\hookrightarrow \bm{b}\mathbb{R}$ is the embedding of $\mathbb{R}$ in $\bm{b}\mathbb{R}$. The scalar product becomes now
$$
\langle \psi_1,\psi_2 \rangle_{\bm{b}\mathbb{R}}=\sum_{x\in \mathbb{R}}\overline{\mathfrak{B}(\psi_1)(x)}\mathfrak{B}(\psi_2)(x)\,,
$$
where it is important to notice that
$$
\mathfrak{B}(\chi_{x_0})=\delta_{x_0}\,,\quad x_0\in \mathbb{R}
$$
so that $\{\chi_x: x\in \mathbb{R}\}$ is an ortonormal basis of $L^2(\bm{b}\mathbb{R},\nu_{\bm{b}\mathbb{R}})$ and any $\psi\in L^2(\bm{b}\mathbb{R},\nu_{\bm{b}\mathbb{R}})$ can be written in a unique form as
$$
\psi=\sum_{x\in \mathbb{R}} \Psi(x) \,\chi_x\,,\quad\textrm{ where }\, \Psi(x)=\mathfrak{B}(\psi)(x)\,,\quad \Psi\in \ell^2(\mathbb{R}).
$$
Equivalently, given $\psi\in AP(\mathbb{R},\mathbb{C})$, its  Fourier-Bohr transform
$$
\mathfrak{B}(\psi)(x):=\langle \phi_x,\psi \rangle_{AP}=\lim_{T\rightarrow \infty}\frac{1}{2T}\int_{-T}^T e^{-ixy} \psi(y) \mathrm{d}y\,.
$$
belongs to $\ell^2(\mathbb{R})$ and  $\{\phi_x: x\in \mathbb{R}\}$ is an ortonormal basis of $AP(\mathbb{R},\mathbb{C})$.

\section{Polymer representations for the Weyl algebra}{\label{PolyWeyl}}
The Weyl algebra, codified by the equation
$$
U(p)V(q)=e^{-ipq}V(q)U(p)\,,\quad \forall q,p\in \mathbb{R}\,,
$$
admits two unitarily inequivalent irreducible representations on $\ell^2(\mathbb{R})$ that we refer to as position  and momentum representations, respectively \cite{Halvorson}:

\begin{itemize}

\item[(\textbf{PR})] \textbf{Position representation} of the Weyl algebra on $\ell^2(\mathbb{R})$: The one-parameter families  of
unitary operators  $\{\bm{U}(p):p\in \mathbb{R}\}$ and $\{\bm{V}(q):q\in \mathbb{R}\}$ defined through
$$
\bm{U}(p)\delta_{q_0}:= e^{ip q_0}\delta_{q_0}\,,\quad \bm{V}(q)\delta_{q_0}:=\delta_{q_0-q}\,,\quad \forall p,q,q_0\in \mathbb{R}\,,
$$
satisfy
$$
\bm{U}(p)\bm{V}(q)=e^{-ipq}\bm{V}(q)\bm{U}(p)\,,\quad \forall q,p\in \mathbb{R}\,.
$$
In this representation,  $q\mapsto \bm{V}(q)$ is not weakly continuous and there is no momentum operator. On the other hand, the map $p\mapsto \bm{U}(p)$ is weakly continuous and there is a position self-adjoint  unbounded operator  $\bm{Q}:=-i\bm{U}'(0)$, satisfying
$$
\bm{Q}\delta_q=q \delta_q\,,
$$
with domain
$$
\mathcal{D}(\bm{Q}):=\{\Psi\in\ell^2(\mathbb{R})\,:\, \sum_{q\in \mathbb{R}}q^2|\Psi(q)|^2<\infty\}\,.
$$

\item[(\textbf{MR})] \textbf{Momentum representation} of the Weyl algebra on $\ell^2(\mathbb{R})$: The one-parameter families  of
unitary operators  $\{\bm{U}(p):p\in \mathbb{R}\}$ and $\{\bm{V}(q):q\in \mathbb{R}\}$ defined through
$$
\bm{U}(p)\delta_{p_0}:= \delta_{p_0+p}\,,\quad \bm{V}(q)\delta_{p_0}:=e^{iqp_0}\delta_{p_0}\,,\quad \forall q,p,p_0\in \mathbb{R}\,,
$$
satisfy
$$
\bm{U}(p)\bm{V}(q)=e^{-ipq}\bm{V}(q)\bm{U}(p)\,,\quad \forall q,p\in \mathbb{R}\,.
$$
In this representation, $p\mapsto \bm{U}(p)$ is not weakly continuous and there is no position operator. On the other hand, the map $q\mapsto \bm{V}(q)$ is weakly continuous and there is a momentum  self-adjoint  unbounded operator   $\bm{P}:=-i\bm{V}'(0)$, satisfying
$$
\bm{P}\delta_p=p \delta_p\,,
$$
with domain
$$
\mathcal{D}(\bm{P}):=\{\Psi\in\ell^2(\mathbb{R})\,:\, \sum_{p\in \mathbb{R}}p^2|\Psi(p)|^2<\infty\}\,.
$$
\end{itemize}

By using the Fourier-Bohr transform, both representations can be implemented in $L^2(\bm{b}\mathbb{R},\nu_{\bm{b}\mathbb{R}})$. In particular, if we use $L^2(\bm{b}\mathbb{R},\nu_{\bm{b}\mathbb{R}})$ as the carrier of the position representation, the operator $\bm{Q}$ is just a derivative operator and $\bm{Q}^2=-\bm{\Delta}$, where $\bm{\Delta}$ is the Laplace operator. On the other hand, in the momentum representation, the operator $\bm{P}$ is just a derivative operator and $\bm{P}^2=-\bm{\Delta}$.

As we will see later when we study the spectra of the polymerized harmonic oscillator Hamiltonians, the eigenvalue equations can be equivalently written as \textit{difference equations} in $\ell^2(\mathbb{R})$ or as \textit{differential equations} in $L^2(\bm{b}\mathbb{R},\nu_{\bm{b}\mathbb{R}})$ with no reason to prefer, \textit{a priori}, one writing over the other. This is in exact analogy with the relationship between the matrix mechanics of Heisenberg and the wave mechanics of Schr\"{o}dinger.

\section{Polymerized Hamiltonians: general results on the spectra}{\label{spectra}}

Let $W:\mathbb{R}\rightarrow \mathbb{R}$ be a periodic piecewise continuous (in the standard topology for $\mathbb{R}$) function, then $W$ defines, through the multiplication in the algebra of almost periodic functions, a bounded operator $\bm{W}$ in $L^2(\bm{b}\mathbb{R},\nu_{\bm{b}\mathbb{R}})$ (equivalently in $AP(\mathbb{R},\mathbb{C})$). The operator
\begin{equation}
\bm{H}=-\bm{\Delta} +\bm{W}\label{operator}
\end{equation}
is a self-adjoint  unbounded  operator with domain
$$
\mathcal{D}(\bm{H})=\mathcal{D}(\bm{\Delta})=\{\psi\in L^2(\bm{b}\mathbb{R},\nu_{\bm{b}\mathbb{R}})\,:\, \sum_{x\in \mathbb{R}} x^4 |\mathfrak{B}(\psi)(x)|^2<\infty\}
$$

The crucial result that allows us to obtain the spectrum of (\ref{operator}) in a straightforward way can be found in \cite{Chojnacki1986,Burnat1964,Chojnacki1991,Krupa} where it is shown that the operator (\ref{operator})  has only \textit{pure point spectrum} that coincides with the spectrum of the Schr\"odinger operator $H=-\Delta+W$,  with periodic potential $W$, defined on $L^2(\mathbb{R},\mu_\mathbb{R})$. We can then use the following well-known theorem \cite{ReedSimon} (that encapsulates the Bloch theorem of the physicists or the Floquet theory of the mathematicians):

\bigskip

\noindent \textbf{Theorem:} \textit{Let $W$ be a piecewise continuous function of period $2\pi$. Let $H=-\Delta+W$ on $L^2(\mathbb{R},\mu_{\mathbb{R}})$. Let $E_1(0)$, $E_2(0),\ldots$ be the eigenvalues of the corresponding operator on $(0,2\pi)$ with periodic boundary conditions and let $E_1(\pi)$, $E_2(\pi),\ldots$ be the eigenvalues with antiperiodic boundary conditions. Let
$$
E^L_n=\left\{\begin{array}{ll}E_n(0)&n\,\,\mathrm{odd}\\E_n(\pi)&n\,\,\mathrm{even}\end{array}\right.
\hspace*{1cm}E^R_n=\left\{\begin{array}{ll}E_n(\pi)&n\,\,\mathrm{odd}\\E_n(0)&n\,\,\mathrm{even}\end{array}\right.\,.
$$
Then $H$ has purely absolutely continuous spectrum and
$$
\sigma(H)=\sigma_{\mathrm{ac}}(H)=\bigcup_{n=1}^\infty [E^L_n,E^R_n]\,.
$$}
Hence we conclude that
\begin{equation}
\sigma(\bm{H})=\sigma_{\mathrm{pp}}(\bm{H})=\sigma(H)=\sigma_{\mathrm{ac}}(H)=\bigcup_{n=1}^\infty [E^L_n,E^R_n]\,.
\label{bands}
\end{equation}
Moreover, the eigenvalues of $\bm{H}$ are, at most, doubly degenerate \cite{Chojnacki1991}. Associated with the bands (\ref{bands}) in the energy spectrum there are spectral gaps of forbidden energies
$$
(-\infty, E^L_1)\cup (E^R_1,E^L_2)\cup(E^R_2,E^L_3)\cup\cdots
$$
Notice that some of the gaps will be absent if the edges $E^R_n$ and $E^L_{n+1}$ of two consecutive  bands coincide. This characterization of the spectrum as consisting of a set of bands is the main result of the paper. We would like to mention here that this fact is somehow implicit in \cite{AFW} (see section 3.2.) although, to our knowledge, the structure of the spectrum of the Hamiltonian as explained above has not be made explicit in the literature.

It is not difficult to understand the previous result from an intuitive point of view. Indeed, the almost periodic solutions that are found by solving the eigenvalue equation (in its differential form) \textit{are not normalizable} in the standard $L^2(\mathbb{R},\mu_\mathbb{R})$ sense but \textit{they are } when the scalar product in the space of almost periodic functions is used. Although it is true that strictly periodic (or anti-periodic) solutions solve the preceding equation there is a continuum of solutions parametrized by a real number --this is the usual statement that underlies the Bloch-Floquet theorem. Indeed, this theorem shows that the non-normalizable ``eigenfunctions'' $\varphi$ of $H=-\Delta+W$ satisfy
\begin{equation}
\varphi(x+2\pi)=e^{2\pi k i}\varphi(x)\,,\quad \forall x\in \mathbb{R}\,,\label{k}
\end{equation}
for some real number $k$ --that can be restricted to lie in  $k\in [0,1)$ or, equivalently, in $k\in[-1/2,1/2)$. The choice  $k=0$ leads to the $L^2(\bm{b}\mathbb{R},\nu_{\bm{b}\mathbb{R}})$  eigenvectors corresponding to ``half'' of the boundary points of the band spectrum  ($2\pi$-periodic solutions, that determine either $E_{2n+1}^L$ or $E^R_{2n}$). The choice  $k=1/2$ leads to  the remaining  boundary points ($2\pi$-antiperiodic solutions, that determine either $E_{2n}^L$ or $E^R_{2n+1}$). The other choices for $k$ give rise to the eigenfunctions corresponding to the rest of the points in the energy bands.

It is straightforward to show that if $\varphi$ satisfies (\ref{k}) it is possible to write
$$
\varphi(x)=e^{ik x } f(x)
$$
for some  $2\pi$-periodic function $f$. Then, it is  obvious that its Fourier-Bohr transform
$$
\mathfrak{B}(\varphi)(x)=\lim_{N\rightarrow\infty}\frac{1}{2\pi N}\int_{- N\pi}^{N\pi} e^{-ix y} \varphi(y) \,\mathrm{d}y=\lim_{N\rightarrow\infty}\frac{1}{2\pi N}\int_{- N\pi}^{ N\pi} e^{-ix y} e^{i k y } f(y)\,\mathrm{d}y
$$
vanishes when
$$
x \not \equiv k  \quad (\mathrm{mod}\,\, 1)\,.
$$
In other words, if $\varphi=\varphi_k$ satisfies (\ref{k}) for some $k$, the set $\{x\in \mathbb{R}\,:\, \mathfrak{B}(\varphi_k)(x)\neq 0\}$ is a subset of the regular lattice $\gamma_k=\{k +n\,:\, n\in \mathbb{Z}\}\subset \mathbb{R}$. As we will see in the next section, this fact has a neat consequence when the eigenvalue equation for the polymer Hamiltonian $\bm H$ is written in $\ell^2(\mathbb{R})$: as elements of $\ell^2(\mathbb{R})$ the eigenfunctions have their support on \textit{regular}, \textit{evenly spaced}, lattices of the previous form. If $k_1\not\equiv k_2\,\, (\mathrm{mod}\,\, 1)$ the lattices $\gamma_{k_1}$ and $\gamma_{k_2}$ are disjoint and, hence, the states $\varphi_{k_1}$ and $\varphi_{k_1}$ are orthogonal. As a final comment we would like to point out here that it is simply not true that the eigenvalue problem for a polymer Hamiltonian $\bm{H}=-\bm{\Delta}+\bm{W}$ is equivalent to the one corresponding to a particle on a unit circle, whose solutions satisfies $\varphi(x+2\pi)=e^{2\pi k i}\varphi(x)$ only for $k=0$.

\section{Different representations for the polymerized harmonic oscillator}{\label{examples}}

The Hamiltonian for the classical harmonic oscillator, written in terms of dimensionless variables $q$ and $p$, has the form
$$
H(q,p)=\frac{\hbar^2}{2m\ell^2} p^2+\frac{m\ell^2 \omega^2}{2} q^2\,,
$$
where $\ell$ has dimension of  length, $m$ has dimension of mass, $\omega$ is the frequency of the harmonic oscillator and $\hbar$ is the  reduced Planck constant. The main difference between the quantizations for the harmonic oscillators in the standard Schr\"{o}dinger representations of the Weyl algebra and the polymer  position or momentum representations is the fact that the Hamiltonian in the latter cases cannot be represented because it is not possible to simultaneously quantize the $q^2$ and the $p^2$ parts of the Hamiltonian.  In order to built the quantum Hamiltonian operator let us work in the position representation. The first step \cite{AFW} (see also \cite{CV}) is to introduce an arbitrary, but fixed, parameter $q_0>0$ (that defines a length scale $q_0\ell$) and consider the following one-parameter family of operators
\begin{equation}
\bm{H}(q_0):=\frac{\hbar^2}{2m(2q_0\ell)^2}\Big( 2 - \bm{V}(2q_0)-\bm{V}(-2q_0)\Big)+ \frac{m\ell^2 \omega^2}{2} \bm{Q}^2.\label{H1}
\end{equation}
By defining the self-adjoint operators
$$
\bm{P}(q_0):=\frac{i}{2q_0}\left(\bm{V}(q_0)-\bm{V}(-q_0)\right)
$$
these Hamiltonians can be written as
$$
\bm{H}(q_0):=\frac{\hbar^2}{2m\ell^2}\bm{P}(q_0)^2+ \frac{m\ell^2 \omega^2}{2} \bm{Q}^2
$$
and, hence, it is straightforward to see that for every value of $q_0>0$ these are non-negative operators in $\ell^2(\mathbb{R})$ with the same domain as $-\mathbf{\Delta}$.

Alternatively, in the momentum representation, the quantum harmonic oscillator Hamiltonian can be modeled by
\begin{equation}
\bm{H}(p_0):=\frac{\hbar^2}{2m\ell^2}\bm{P}^2+ \frac{m\ell^2 \omega^2}{2 (2p_0)^2}\Big( 2 - \bm{U}(2p_0)-\bm{U}(-2p_0)\Big),\label{H2}
\end{equation}
where $p_0>0$ introduces a momentum scale $p_0\hbar/\ell$. If the momentum and length scales defined by the parameter $q_0$ and $p_0$ satisfy
$$
\frac{\hbar}{\ell} p_0 =m\omega q_0 \ell
$$
the spectra of (\ref{H1}) and (\ref{H2}) coincide so we will use (\ref{H1}) in the following. (Notice, however, that the momentum representation is the one used in many polymer quantum field theories \cite{L1,L2,L3}). A comment is in order here: the approximation that is used in (\ref{H1}) to make sense of  $p^2$ in the polymer position representation can be seen as just a simple and natural choice among a large family of possible ones (as mentioned in \cite{CV}); another such --and natural-- choice will be considered below.

We are interested in studying the spectrum of the Hamiltonian (\ref{H1}) as a function of the parameter $q_0$. To this end we will write the eigenvalue equation both in the $\ell^2(\mathbb{R})$ and $L^2(\bm{b}\mathbb{R},\nu_{\bm{b}\mathbb{R}})$ position representations. The eigenvalue equation in the $\ell^2(\mathbb{R})$  position representation is
$$
\frac{\hbar^2}{2m(2q_0\ell)^2}\Big( 2 \Psi(q) - \Psi(q+2q_0)-\Psi(q-2q_0)\Big)+ \frac{m\ell^2 \omega^2}{2} q^2 \Psi(q)=E \Psi(q)\,,\,\,\forall q\in \mathbb{R}
$$
where the solutions $\Psi\neq 0$ must satisfy the normalizability condition:
$$
\sum_{q\in \mathbb{R}}|\Psi(q)|^2<\infty\,,\quad \sum_{q\in \mathbb{R}}q^4|\Psi(q)|^2<\infty\,,
$$
and the non-negativity of the Hamiltonian guarantees that the only possible eigenvalues $E$ must satisfy $E\geq 0$. This is a difference equation that establishes a relationship between the values of $\Psi$ at three different points $q$, $q+2q_0$ and $q-2q_0$.  Actually, as mentioned at the end of section \ref{spectra} (see also \cite{AFW}), it is possible to solve this difference equation  by looking for solutions $\Psi_k$, where $k\in[0,1)$,  for which $\{q\in \mathbb{R}\,:\,\Psi_k(q)\neq 0\}$ is a subset of the regular lattice
$$
2q_0\cdot \gamma_{k}:=\{2q_0 k+ 2q_0n\,:\, n\in \mathbb{Z}\}\,.
$$
On the other hand, the eigenvalue equation in $L^2(\bm{b}\mathbb{R},\nu_{\bm{b}\mathbb{R}})$ is the differential equation
\begin{equation}
\psi''(p)+\left(\frac{2E}{m\ell^2 \omega^2}-\frac{\hbar^2}{ m^2\ell^4\omega^2q_0^2}\sin^2(q_0 p)\right)\psi(p)=0\label{eigenvalues}
\end{equation}
that must be solved for those $\psi\in L^2(\bm{b}\mathbb{R},\nu_{\bm{b}\mathbb{R}})$ belonging to the domain of the operator $\bm{Q}^2=-\bm{\Delta}$.
By performing the change of variables
$$
x=2q_0 p
$$
and defining $\varphi(x)$ such that
$$
\psi(p)=\varphi(2q_0 p)\,,\quad \psi''(p)=4q_0^2\varphi''(2q_0 p)\,,
$$
the eigenvalue equation (\ref{eigenvalues}) can be rewritten in the form of the familiar Mathieu equation
\begin{equation}
\varphi''(x)+\left(\frac{E}{2m\omega^2(q_0\ell )^2}-\frac{1}{2}\left(\frac{\hbar}{2m \omega(q_0\ell)^2 }\right)^2(1-\cos x)\right)\varphi(x)=0\label{Mathieu}
\end{equation}
for $\varphi\in L^2(\bm{b}\mathbb{R},\nu_{\bm{b}\mathbb{R}})$.

Notice that both equations (\ref{eigenvalues}) and (\ref{Mathieu}) can be rewritten as an eigenvalue problem of the form
$$
\big(-\bm{\Delta}+\bm{W}(q_0)\big)\psi = \lambda \psi\,,\quad \psi\in L^2(\bm{b}\mathbb{R},\nu_{\bm{b}\mathbb{R}})\,.
$$
In particular, for equation  (\ref{eigenvalues}) we have
$$\lambda=\frac{2E}{m\ell^2 \omega^2}$$
and the bounded operators $\bm{W}(q_0)$ are defined by the $\pi/q_0$-periodic functions  $W(\cdot|q_0):\mathbb{R}\rightarrow \mathbb{R}$:
$$
W(p|q_0)=\frac{\hbar^2}{ m^2\ell^4\omega^2q_0^2}\sin^2(q_0 p)\,.
$$
On the other hand, equation (\ref{Mathieu}) is written in terms of  the $2\pi$-periodic potentials $w(\cdot|q_0):\mathbb{R}\rightarrow \mathbb{R}$ defined by
$$
w(x|q_0):=\frac{1}{2}\left(\frac{\hbar}{2m \omega(q_0\ell)^2 }\right)^2\big(1-\cos x\big)\,.
$$
It is very important to emphasize at this point that, as stated in section \ref{spectra}, in order to obtain the energy spectrum, we must find the band spectrum of the corresponding Sch\"odinger operators in $L^2(\mathbb{R},\mu_\mathbb{R})$. We would like to mention here that the equation (\ref{Mathieu}),  considered as an eigenvalue problem in the separable Hilbert space $L^2(\mathbb{S}^1,\mu_{\mathbb{S}^1})$, where $\mu_{\mathbb{S}^1}$ is the Haar measure for $\mathbb{S}^1$, was proposed in \cite{Chalbaud} to study the quantum  Harmonic oscillator on a lattice. Notice that the non-trivial solutions $\varphi$ to equation (\ref{Mathieu}) that satisfy  $\varphi\in L^2(0,2\pi)$, $\varphi(2\pi)=\varphi(0)$, and $\varphi'(2\pi)=\varphi'(0)$ are just the solutions to (\ref{Mathieu}) in $L^2(\mathbb{S}^1,\mu_{\mathbb{S}^1})$. These are also relevant when (\ref{Mathieu}) is considered in $L^2(\bm{b}\mathbb{R},\nu_{\bm{b}\mathbb{R}})$  because they provide the eigenvectors corresponding to certain boundary points of the band spectrum  (either $E_{2n+1}^L$ or $E^R_{2n}$). The antiperiodic solutions $\varphi\in L^2(0,2\pi)$, satisfying $\varphi(2\pi)=-\varphi(0)$ and $\varphi'(2\pi)=-\varphi'(0)$, provide the eigenvectors corresponding to the remaining  boundary points of the bands (either $E_{2n}^L$ or $E^R_{2n+1}$). It is important to notice that, in addition to these, there are solutions $\varphi\in L^2(0,2\pi)$ satisfying  $\varphi(2\pi)=e^{2\pi k i }\varphi(0)$ and $\varphi'(2\pi)=e^{2\pi k i}\varphi'(0)$, for any $k\in \mathbb{R}$, that give rise to the eigenfunctions of the rest of the points in energy bands.

The general results of section \ref{spectra} guarantee that
$$
\sigma(\bm{H}(q_0))=\sigma_{\mathrm{pp}}(\bm{H}(q_0))=\bigcup_{n=1}^\infty[E^L_n(q_0),E^R_n(q_0)]\,,\quad E^L_n(q_0)<E^R_n(q_0)\,.
$$
In this case, it is also known \cite{AS} that (see, also \cite{AFW,Husain1})
$$
\lim_{q_0\rightarrow 0}E^R_n(q_0)=\lim_{q_0\rightarrow 0}E^L_n(q_0)=E_n\,,
$$
where $E_n$ in the $n$-th eigenvalue of the $L^2(\mathbb{R},\mu_{\mathbb{R}})$ harmonic oscillator Hamiltonian
$$
H_{\mathrm{HO}}=- \frac{m\ell^2 \omega^2}{2} \Delta + W_{\mathrm{HO}}\,,
$$
where $W_{\mathrm{HO}}:\mathbb{R}\rightarrow \mathbb{R}$ is the quadratic potential
$$
W_{\mathrm{HO}}(p)=\frac{\hbar^2}{ m^2\ell^4\omega^2} p^2\,.
$$
Notice, however, that for a fixed value of $q_0>0$ \textit{all} the gaps between bands are present (see, for example, \cite{ReedSimon}), i.e.
$$
D_n(q_0):= E^L_{n+1}(q_0)-E^R_n(q_0)>0\,,\quad \forall n=1,2,\ldots
$$
Actually, the width of the gaps behaves as \cite{Simon}
$$
D_n(q_0)=\frac{m\omega^2(q_0\ell)^2}{4^{n-1}((n-1)!)^2}\left(\frac{\hbar}{2m\omega (q_0\ell)^2}\right)^{2n} \Big(1+o(1/n^2)\Big)\,.
$$
As we can see they get narrower and narrower as the energy grows. In practice this means that for high energies the spectrum basically consists of a continuum of eigenvalues belonging to the point spectrum (i.e. associated with \textit{normalizable} states). This behaviour is very different from that of the ordinary quantum harmonic oscillator.

A comment is in order now. Although the simplest way to model the standard harmonic oscillator Hamiltonian in the polymer position representation
is by approximating $p^2$ as
$$
\bm{P}(q_0)^2=\frac{m^2\ell^4\omega^2}{\hbar} \bm{W}(q_0)\,,\quad q_0>0\,,
$$
there is, in fact, a huge ambiguity here (as has been realized by other authors \cite{CV}). In particular,  in the position representation supported by $L^2(\bm{b}\mathbb{R},\nu_{\bm{b}\mathbb{R}})$, one can actually use, for example, any $\pi/q_0$-periodic function of $W(\cdot|q_0):\mathbb{R}\rightarrow \mathbb{R}$ such that
$$
\lim_{q_0\rightarrow 0}W(p|q_0)= \frac{\hbar^2}{m^2\ell^4\omega^2} p^2\quad \textrm{(pointwise)}\,.
$$
 A very simple such choice is provided by the $\pi/q_0$-periodic extension of a purely quadratic (pq) potential
$$
W_{\mathrm{pq}}(p|q_0):=\frac{\hbar^2}{ m^2\ell^4\omega^2}p^2\,,\quad p\in \big[-\frac{\pi}{2q_0},\frac{\pi}{2q_0}\big]\,.
$$
The band structure in this case is qualitatively equal to the one described for the Mathieu equation. As can be seen in the figure,  the first bands closely correspond to the usual eigenvalues of the harmonic oscillator and are very narrow. On the other hand it can be readily seen that as the energy grows the bands widen and the gaps become more and more narrow, in agreement with the behavior described above in the case of the Mathieu approximation.

An interesting question that arises at this point is whether all these possible extensions behave as the standard quantum harmonic oscillator in the $q_0\rightarrow 0$ limit or, more precisely, how do all the different spectra compare in this limit. This question is made even more relevant by the fact that there are periodic potentials (leading, for example, to the so called Lam\'e equation) that give rise only to a finite number of bands in the spectrum. If such potentials could be used to approximate a periodic extension of a purely quadratic one with arbitrary large period it is difficult to understand to what extent the spectrum of the ordinary quantum harmonic oscillator could be recovered. A partial answer can be given by considering pairs of Hamiltonians that are close to each other and using perturbation theory. If one takes $\bm{H}_0=-\bm{\Delta} +\bm{W}_0$, $\bm{H}_1=-\bm{\Delta} +\bm{W}_1$ where $\bm{W}_1=\bm{W}_0+\bm{V}$ and $||\bm{V}||<\epsilon$ then $\bm{H}_1=\bm{H}_0+\bm{V}$ and the eigenvalues of $\bm{H}_1$ are approximated by those of $\bm{H}_0$ with $O(\epsilon)$ corrections. Notice, however, that if there is no need to recover $p^2$ for all $p\in \mathbb{R}$,  \textit{any potential} term that has a quadratic minimum at $p=0$ of the same type as the one in the potential that appears in the Mathieu equation provides a perfectly valid approximation for the harmonic oscillator Hamiltonian at small energies. An interesting example in this respect is provided by periodic potentials defined by Jacobian elliptic functions, in particular those leading to the Lam\'e equation \cite{Finkel}. It is known that, in those cases the spectrum consists of a continuum of positive eigenvalues with a finite number of gaps (or, equivalently, on a finite number of bounded bands and an unbounded band). These examples show that one has to be careful when extrapolating the form of the spectrum from the band structure of the polymerized quantum Hamiltonians for the harmonic oscillator.

\begin{figure}
\begin{center}
\includegraphics[width=6.3cm]{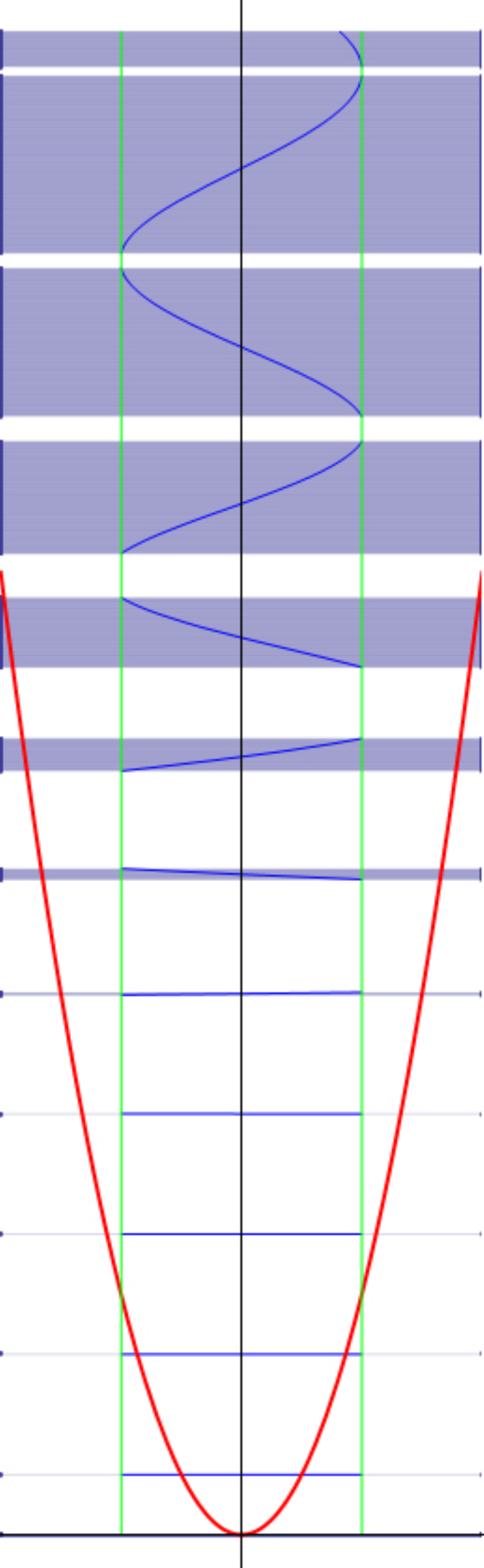}
\end{center}
{\parbox{14.4cm}{\hspace*{-2cm}\caption{Band structure for $H=-\Delta+W_{\mathrm{pq}}$ where $W_{\mathrm{pq}}$ is the periodic potential satisfying  $W_{\mathrm{pq}}(x)=x^2$ for $x\in[-4,4]$. This corresponds to choosing $q_0=\pi/8$ and $\ell=\hbar=m=\omega=1$. We have plotted the potential in one period, the bands that constitute the spectrum $\sigma(H)$, and the trace of the matrix $M(E)$ that determines the position of the bands as those values of the energy for which $\mathrm{Tr} \, M(E)=\pm 2$ (see \cite{ReedSimon}). The narrow lowest energy bands closely correspond to the lowest quantum harmonic oscillator eigenvalues.}}}
\end{figure}

\section{Statistical mechanics of the polymerized harmonic oscillator}{\label{statistical}}

We briefly discuss in this section the statistical mechanics of a single particle whose quantum  dynamics is governed by a Hamiltonian $\bm{H}=-\bm{\Delta}+\bm{W}$ belonging to the class (\ref{operator}). We will show that neither the microcanonical nor the canonical ensembles provide a convenient starting point to study the statistical mechanics of the system. This is so because both the statistical (counting) entropy and the partition function for the system are ill-defined.  Notice that, at face value, this would imply that fundamental model systems such as the Einstein crystal would be ill defined if the polymer quantization of the harmonic oscillator is used.

For $E\in \sigma(\bm{H})=\sigma_{\mathrm{pp}}(\bm{H})$, let us denote by $\bm{P}_{\bm{H}}(E)$ the self-adjoint orthogonal projector on the eigenspace $\bm{H}=E$. These projectors give rise to the resolution of the identity
$$
\bm{I}=\sum_{E\in \sigma(\bm{H})} \bm{P}_{\bm{H}}(E)\quad\textrm{ i.e. }\quad \psi=\sum_{E\in \sigma(\bm{H})} \bm{P}_{\bm{H}}(E)\psi\,,\quad \forall \psi\in L^2(\bm{b}\mathbb{R},\nu_{\bm{b}\mathbb{R}}).
$$
The statistical entropy for the system as a function of the energy is given by
$$
\Omega(E):=\mathrm{Tr} \Big(\sum_{\epsilon\in\sigma(\bm{H})\cap (-\infty,E]}\bm{P}_{\bm{H}}(\epsilon) \Big)=\sum_{\epsilon\in\sigma(\bm{H})\cap (-\infty,E]}d(\epsilon)
$$
where $d(E)$ is the degeneracy of the energy eigenvalue $E$. As we mentioned above (see \cite{Chojnacki1991}) $d(E)$ is at most two for generic one dimensional problems described by (\ref{operator}). As it is plainly obvious the continuous sums written above are ill defined owing to the uncountable number of energy eigenstates of the system. Likewise the canonical partition function is also ill-defined. In this case we have to consider (by using the spectral representation theorem)
$$
e^{-\beta \bm{H}}=\sum_{E\in \sigma(\bm{H})} e^{-\beta E} \bm{P}_{\bm{H}}(E)\,,\quad \beta>0\,.
$$
Notice that $e^{-\beta \bm{H}}$ is a bounded operator because
$$
||e^{-\beta \bm{H}}\psi||^2=\sum_{E\in \sigma(\bm{H})} e^{-2\beta E} || \bm{P}_{\bm{H}}(E) \psi||^2\leq \sum_{E\in \sigma(\bm{H})}  || \bm{P}_{\bm{H}}(E) \psi||^2=||\psi||^2\,,\quad \forall\psi\in L^2(\bm{b}\mathbb{R},\nu_{\bm{b}\mathbb{R}})
$$
but it is not a trace class operator,
$$
Z(\beta)=\mathrm{Tr}(e^{-\beta \bm{H}})=\sum_{E\in \sigma(\bm{H})} d(E) e^{-\beta E} =\infty,
$$
because $\#\{E\in \sigma(\bm{H})\,:\, d(E)e^{-\beta E}>0 \}=\# \mathbb{R}$. Hence the thermal density matrix
$$
\bm{\rho}=\frac{e^{-\beta \bm{H}}}{\mathrm{Tr}(e^{-\beta \bm{H}})}
$$
is ill defined.

Notice that the situation that we are facing here is conceptually different from the one corresponding to a standard Schr\"odinger operator $H=-\Delta+W$ in $L^2(\mathbb{R},\mu_{\mathbb{R}})$ for a one-dimensional periodic potential $W$. Indeed, despite the fact that in that case, as discussed in section \ref{spectra}, the spectrum of $H$ is \textit{purely absolutely continuous}, bounded from below, and the operators $e^{-\beta H}$ are trace class  for $\beta>0$. In particular, the microcanonical entropy is given by
$$
\Omega(E)=\int_{\sigma(H)\cap(-\infty,E]}g(\epsilon)\mathrm{d}\epsilon
$$
in terms of $g(E)$, the density of states per energy interval, and the partition function is
$$
Z(\beta)=\int_{\sigma(H)}g(E)e^{-\beta E}\mathrm{d}E\,.
$$
The statistical entropy is obviously well defined, the partition function too (as long as the density of states $g(E)$ does not grow too fast) and we have the standard probabilistic interpretation associated with the canonical ensemble:
$$
\mathrm{Prob}(B|\beta)=\frac{1}{Z(\beta)}\int_B g(E)e^{-\beta E}\mathrm{d}E\,,\quad B\subset \sigma(H)\,.
$$

A possible way out of the situation explained above would be to find a physical criterion to eliminate most of the eigenstates of the Hamiltonian and just leave a countable number of them. Suppose that, for every $n\in \mathbb{N}$, we select one energy eigenvalue in each band, $E_n\in [E_n^L,E_n^R]$, with $E_n<E_{n+1}$ and $\lim_{n\rightarrow\infty}E_n=\infty$. In general, let
$$
\mathcal{E}:=\bigcup_{n=1}^\infty\{E_n\}\,,\quad \bm{P}_{\mathcal{E}}:=\sum_{E\in \mathcal{E}} \bm{P}_{\bm{H}}(E)\,,\quad
\bm{H}_{\mathcal{E}}:= \bm{H} \bm{P}_{\mathcal{E}}=\bm{P}_{\mathcal{E}} \bm{H} \bm{P}_{\mathcal{E}}\,,
$$
then
$$
\Omega_{\mathcal{E}}(E):=\mathrm{Tr} \Big(\sum_{\epsilon\in\mathcal{E}\cap (-\infty,E]}\bm{P}_{\bm{H}}(\epsilon) \Big)=\sum_{\epsilon\in\mathcal{E}\cap (-\infty,E]}d(\epsilon)=\sum_{E_n\leq E} d(E_n)
$$
and
$$
Z_{\mathcal{E}}(\beta)=\mathrm{Tr}(e^{-\beta \bm{H}_\mathcal{E}})=\sum_{E\in \mathcal{E}} d(E) e^{-\beta E}=\sum_{n=1}^\infty d(E_n) e^{-\beta E_n}< \infty\,,\quad \forall \beta>0\,,
$$
are well defined.

We would like to mention here that, within the context of LQG, it is customary to talk about superselected sectors in the the full gravity Hilbert space. A mechanism that may be at work in this setting is the selection of a separable physical subspace of the full non-separable Hilbert space in the process of implementing the quantum constraints \textit{\`a la Dirac}. This has been checked in detail in the context of the polymer quantization of parameterized field theories by Laddha and Varadarajan \cite{Varadarajan1,Varadarajan2}.

Superselection rules may be the ingredient needed to avoid some of the statistical problems mentioned above. In particular,  given  $k\in [0,1)$, let us consider \cite{AFW} the  subspace $\mathcal{H}_k\subset L^2(\bm{b}\mathbb{R},\nu_{\bm{b}\mathbb{R}})$ that is generated by the energy eigenfunctions satisfying $\varphi(x+2\pi)=e^{2\pi k i}\varphi(x)$. The Fourier-Bohr transform of any of such eigenvectors fulfils
$$
\{x\in \mathbb{R}\,:\, \mathfrak{B}(\varphi)(x)\neq 0\}\subset \gamma_k=\{k+n\,:\, n\in \mathbb{Z}\}
$$
and it is possible to prove that each subspace  $\mathcal{H}_k$  is ``superselected'' in the sense that it its not only trivially invariant under the action of $\bm{H}$ but it is also invariant under both the action of the corresponding self-adjoint position operator and that of an appropriately chosen self-adjoint operator that approximates the momentum in the polymer position representation \cite{AFW}. (However, it is important to notice that  it is possible  to approximate the momentum is such a way that the subspaces  superselected in that way become  $\mathcal{H}_k\oplus \mathcal{H}_{1/2-k}$, for $k\in[0,1/2)$, or even more general combinations.). The canonical ensemble defined by the restriction of the harmonic oscillator Hamiltonian to one of such $\mathcal{H}_k$-sectors has been considered in  \cite{Chalbaud}. However, notice that, as mentioned in standard references on superselection rules (see, for instance, \cite{Wightman}), it is perfectly acceptable to have density matrices involving states in different superselected sectors so, in principle, one should not exclude statistical ensembles including such states. The existence of superselected subspaces in a physical Hilbert space, rather than precluding the possibility of considering non-trivial linear combinations of states in different sectors, entails the \textit{physical impossibility} of measuring the relative phases between them. This, in turn, means that such states are physically realizable only as density matrices. In any case, unlike the situation in LQG, in the simple quantum mechanical model that we have considered in the paper, this recourse to superselection rules does not seem to be justified. This is so because, if its sole purpose is to render the Hilbert space separable by restricting the quantum dynamics to a Hilbert subspace consisting of an orthonormal sum of eigespaces of the Hamiltonian, there is no point in introducing a polymer quantization to begin with. Standard quantum mechanics would suffice.

\section{Final comments and conclusions}{\label{conclusions}}

The main result discussed in the paper is the fact that, generically, the spectrum of the polymer Hamiltonians for the harmonic oscillator consists of \textit{bands} similar to the ones appearing in the study of periodic potentials in ordinary quantum mechanics. This means that the limit in which the length scale defined by $q_0$ becomes very small should be handled with proper care.  Although there is evidence supporting that the energy eigenvalues for the standard quantum harmonic oscillator are recovered in this limit, important features of the spectrum differ in significant ways, the most important one being the fact that the effective degeneracy of the energy eigenvalues becomes infinite (an \textit{uncountable infinite}, in fact).

The underlying reason for this is easy to understand: the differential symbol for the eigenvalue equation in the position representation is the standard one for a particle in a periodic potential. The fact that the scalar product is not the standard one renders the solutions to the equation (for suitable values of the energy) normalizable (whereas they are not if the standard scalar product in $L^2(\mathbb{R},\mu_{\mathbb{R}})$ is used). In this sense the periodic and antiperiodic solutions that are mentioned in the literature are just a small subset of all the (almost-periodic) solutions permitted by the Bloch theorem. Notice, by the way, that it is not correct to think of the polymerized Hamiltonians as corresponding to a particle in a circle with a potential with a second order minimum (where only periodic --and not antiperiodic-- ``boundary conditions'' would make sense). The correct point of view is to think about such systems as consisting of a particle in a periodic potential with wave functions defined in the Hilbert space of almost periodic functions.

There are a number of results in the literature regarding the band structure for periodic potentials \cite{ReedSimon} and proofs of specific properties of the spectrum for particular choices. The details of how the limit in which the parameter $q_0$ goes to zero works are subtle. As has been shown in the examples of section \ref{examples} the number of bands appearing in the spectrum for one-dimensional polymer Hamiltonians of the type that we have discussed here may be finite (as in the case of the Lam\'e potentials, see \cite{Finkel} and references therein) or infinite as it happens in the case of the usual Hamiltonian leading to the Mathieu equation.

The quantization of many free theories relies heavily on the quantum harmonic oscillator as these systems can be thought of as consisting of an infinite number of such systems. A possible approach to the quantization of these models \cite{Husain1} is based on the construction of a Fock space on the polymer Hilbert space considered here (see \cite{L1,L2,L3,DDL} for different approaches) . In principle it is to be expected that the band structure that we have discussed may play an important role and have observable consequences. The easiest way to explore those is by looking to 2-point functions and, very especially, to field commutators. It is interesting, in particular, to understand if the polymerization changes the microcausality of the system.

The impossibility of defining a partition function for the polymerized models discussed here by generalizing the standard expressions in quantum statistical mechanics (involving traces and trace-class operators) is, to say the least, disturbing and we do not see an obvious way out in this setting. Notice, however, that this negative conclusion should not be naively carried over to LQG. First of all, as mentioned above, the final physical Hilbert space of LQG can very well be separable despite the fact that the kinematical Hilbert space is not thus avoiding any difficulties of the type explained above. Also, at least from the point of view of gravity, the fact that one cannot define the energy in a natural way leads naturally to the consideration of ensembles associated with operators different from the Hamiltonian, in particular the area operator (with its discrete spectrum and effectively finite degeneracies) that is customarily used, for example, to discuss black hole entropy in LQG \cite{ABCK,Nos,EPN}. This change of ensemble may be what is needed to have a sensible framework for the statistical mechanics of quantum gravitational systems \cite{Krasnov,BV}.

To end we would like to mention that, although it is not straightforward to extend the results of the present paper to loop quantum cosmology (and, actually, they may not be directly applicable) we think that it may be instructive to look at the differential version of the difference equations that are usually employed to discuss the quantum dynamics of the early universe in LQC. In that context the argued presence of superselection rules (that remove huge sectors from the space of quantum states) seems to sidestep some of the difficulties that crop up in the simple model provided by the polymerized harmonic oscillator. We think, however, that the issues that we have raised in the paper may be relevant for more advanced models incorporating more degrees of freedom.

\acknowledgements

We would like to thank A. Corichi, J. Lewandowski, T. Paw{\l}owski, P. Singh and M. Varadarajan for their valuable comments. This work has been supported by the Spanish MICINN research grants FIS2009-11893, FIS2012-34379 and the  Consolider-Ingenio 2010 Program CPAN (CSD2007-00042).

\end{document}